# Mechanical behavior of InP twinning superlattice nanowires

Zhilin Liu[a, b], I. Papadimitriou[b], M. Castillo-Rodríguez[b], C. Wang[b], G. Esteban-Manzanares[b],

Xiaoming Yuan[c, *], H. H. Tan[d], J.M. Molina-Aldareguía[b], J. Llorca[b, e, *]

[a] *State Key Laboratory of High Performance Complex Manufacturing, College of Mechanical and Electrical Engineering, Central South University, 410083, P.R. China*

[b] *IMDEA Materials Institute, C/Eric Kandel 2, 28906, Getafe, Madrid, Spain*

[c] *Hunan Key Laboratory for Supermicrostructure and Ultrafast Process, School of Physics and Electronics, Central South University, Changsha, Hunan 410083, P. R. China*

[d] *Department of Electronic Materials Engineering, Research School of Physics and Engineering, The Australian National University, Canberra, ACT 0200, Australia*

[e] *Department of Materials Science, Polytechnic University of Madrid, E.T.S. de Ingenieros de Caminos, 28040 Madrid, Spain*

Corresponding authors: xiaoming.yuan@csu.edu.cn；  javier.llorca@imdea.org

**Abstract**

Taper-free InP twinning superlattice (TSL) nanowires with an average twin spacing of ~ 13 nm were grown along the zinc-blende close-packed [111] direction using metalorganic vapor phase epitaxy. The mechanical properties and fracture mechanisms of individual InP TSL nanowires in tension were ascertained by means of *in situ* uniaxial tensile tests in a transmission electron microscope. The elastic modulus, failure strain and tensile strength along the [111] direction were determined. No evidence of inelastic deformation mechanisms was found before fracture, which took place in a brittle manner along the twin boundary. The experimental results were supported by molecular dynamics simulations of the tensile deformation of the nanowires that also showed that the fracture of twinned nanowires occurred in the absence of inelastic deformation mechanisms by the propagation of a crack from the nanowire surface along the twin boundary.

**Keywords:** Twinning superlattice, InP nanowires, *In-situ* TEM mechanical test, molecular dynamics, fracture behavior

## 1. Introduction

InP semiconductor nanowires have been shown to carry charge and excitons efficiently [1] and also exhibit very high optical emission efficiency [2]. Compared with other III-V nanowires, the InP nanowire array solar cells have the highest efficiency of 17.8% [3] and show significant potential applications for nanoelectronic and optoelectronic devices [1, 4-6] and biochemical sensors [7]. The electronic quality of semiconductor nanowires mainly depends on two factors, i.e. the stacking faults and the crystal structure [8] and the former has already been studied elsewhere [9, 10]. Twinning superlattice (TSL) nanowires have a long-range periodic order of twin planes, which promotes the formation of minibands through tuning the electronic band structure [11, 12]. Recently, small domain TSLs were reported to form locally in bulk Si epitaxial layers [13], and occasionally in tapered InP [8] and ZnS nanowires [14]. However, taper-free TSL nanowires are rarely obtained. Moreover, it has been well documented that the mechanical strain strongly influences the electric, optical and magnetic properties of nanomaterials [10, 15-19]. For instance, lattice straining is known to enhance the drive current in complementary metal-oxide semiconductors technology by improving the electron and hole mobility [17, 19]. Additionally, nanowires may also be exposed to mechanical loading/unloading during fabrication, packaging and service, which may lead to fracture. Thus, understanding the mechanical behavior of InP nanowires (and their corresponding physical mechanisms) is important to ensure the reliability of nanowire-based devices. Nevertheless, this information is very limited due to the difficulties associated with the mechanical testing of nanowires.

Accurate mechanical characterization of nanowires can be carried out by means of *in situ* tensile tests within a transmission electron microscope (TEM). This technique requires sophisticated sample preparation, manufacturing of nanowires suitable for tension tests, a high-speed charge-coupled device camera, and a loading frame with high displacement/load resolution. Owning to these difficulties, most of available information about the mechanical deformation of nanowires has been achieved through nanocompression [10, 20-22], nanobending [23-25], qualitative nanotension (load was not measured) [26], mechanical clamping [27], and *in situ* straining within the scanning electron microscope [28]. Both ends of nanowires may not be exactly placed on the same plane in these approaches, which leads to misalignments during tensile deformation, introducing spurious effects. Moreover, other testing approaches to determine the elastic modulus based on scanning probe microscopy (SPM) lead to a very large experimental error [29]. For instance, the experimental errors in the elastic moduli of wurtzite- and zinc-blende InP nanowires were of the order of 30 and 100 GPa, respectively [29]. In recent years, several experimental techniques have been developed to characterize the mechanical properties of nanomaterials, including the SPM bending, atomic force microscopy (AFM) bending, nanotensile tests (with different mechanical clamps), dynamic resonance (within the AFM and TEM), nanocompression (with either a flat punch, a wedge indenter, or a pyramidal indenter), nanoscratching and nanolooping tests [30-34]. They have been used to explore the mechanical behavior of metallic nanomaterials, including Cu nanocrystals [21, 27], Mo alloy nanofibers [35], Cu/Nb nanolaminate [36], Pt nanocrystal [26], intermetallic $Ni_2Si$ nanowires [37], lithiated Si nanowires [38], Au nanowires [39], and Ag nanopillars [22]. However, the information available for the uniaxial tensile deformation of single semiconductor nanowires is very limited [40, 41].

In this investigation, the mechanical behavior under uniaxial tension of taper-free InP twinning superlattice nanowires was measured *in situ* within the TEM using a push-to-pull (PTP) device. The tensile stress-strain curve and the fracture mechanisms were determined for the first time. Then, additional insights were derived by comparison of the experimental results with molecular dynamics (MD) simulations. The present work provides three fundamental results: (a) the *in situ* observation of the failure behaviour of InP TSL nanowires, (b) their mechanical properties (i.e. elastic modulus and fracture strength), and (c) the effect of twins on the tensile fracture mechanism. This information is important for the design of reliable InP nanowire-based nanodevices with respect to mechanical deformation.

## 2. Material and experimental procedures

### 2.1. Fabrication of InP superlattice nanowires

The InP TSL nanowires were grown by Au-seeded vapor-liquid-solid mechanism in a metalorganic vapor phase epitaxy reactor (MOVPE, Aixtron 200/4) at 100 mbar and with a total gas flow rate of 15 litres per minute [42]. Trimethylindium (TMIn) and phosphine ($PH_3$) were used as precursors for In and P, respectively. Before growth, InP(111)B substrates were immersed in poly-L-lysine (PLL) solution for one minute before rinsing with deionized water. Then, Au droplets with a diameter of 30 nm were deposited on the treated InP(111)B substrates for 30 seconds, followed by cleaning using deionized water. After Au droplet functionalization, the substrates were loaded into the reactor and the InP nanowires with TSL structure were grown by using the reported approach [42]. Stable nanowire growth was initiated by pre-nucleation of the nanowires at 450 °C for 5 minutes. Thereafter, the growth temperature was ramped up to 600 ºC for one hour, promoting the formation of the TSL structure. The TMIn flow rate and V/III ratio were kept at $0.8 \times 10^{-5}$ mol/min and 1116 during growth, respectively. Afterwards, the InP nanowires were cooled down to room temperature in a protective atmosphere of $PH_3$ gas. The morphology and crystal structure of the as-grown nanowires were characterized by field-emission scanning electron microscopy (FE-SEM, Zeiss Ultraplus) at 5 kV and by scanning transmission electron microscopy (STEM, JEOL2100F) at 200 kV. The SEM micrographs are shown in Figures 1a,b.

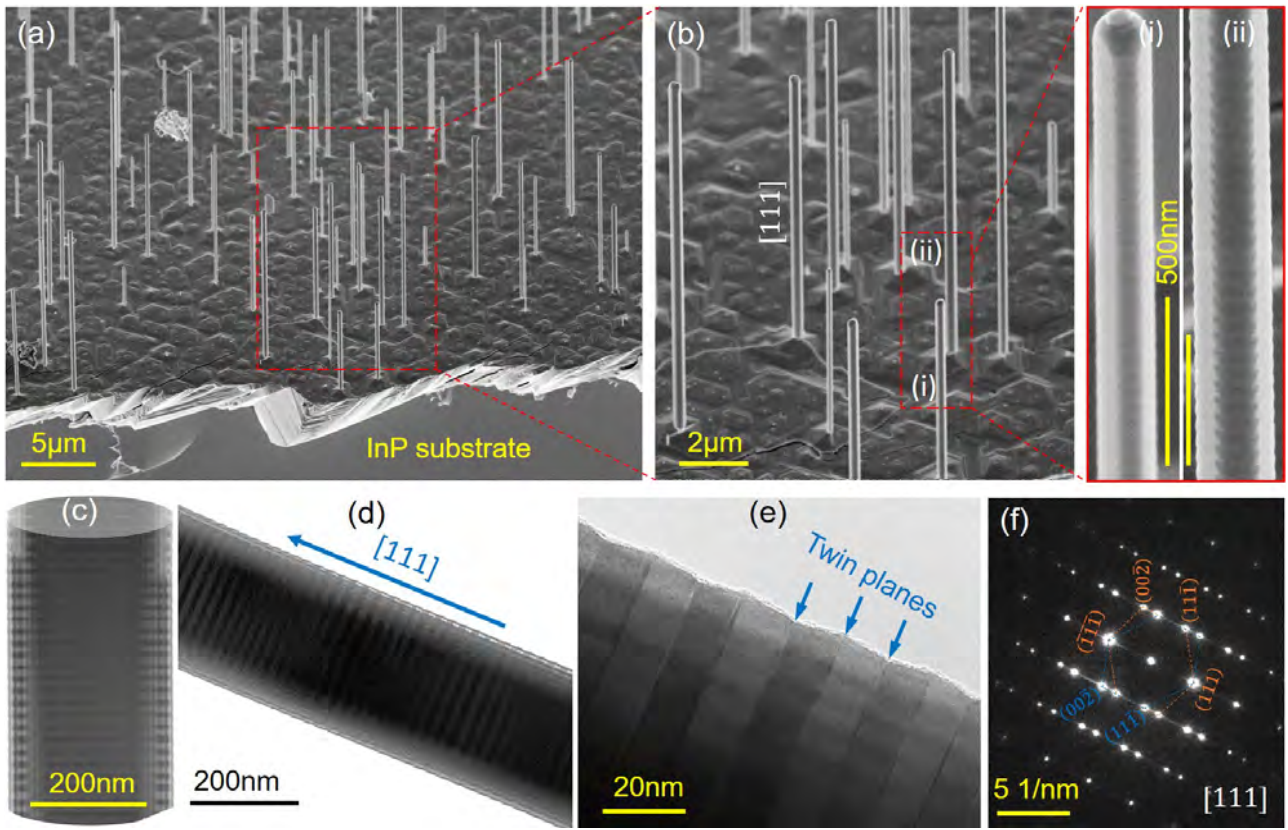

Figure 1. Microstructural characteristics of taper-free InP TSL nanowires. (a, b) SEM images of InP superlattice nanowires grown on InP(111)B substrate. (c, d) TEM images showing the taper-free InP superlattice nanowires with an average twin distance of ~ 13 nm. (e, f) High-resolution TEM image and the corresponding selected area electron diffraction (SAED) pattern of a single InP superlattice nanowire.

### 2.2. In situ TEM mechanical testing

Mechanical deformation of InP TSL nanowires was carried out in a TEM (Talos F200X, FEI, USA) operated at 200 keV with controlled current to reduce beam damage on the nanowires. The *in situ* mechanical deformation was applied through a diamond flat punch (SN: 5-2039-112) coupled with a push-to-pull (PTP) loading device, as shown in Figures 2a,b. The flat punch is actuated using a PI95 TEM PicoIndenter holder (from Hysitron Inc., USA) [21, 36, 41, 43]. Using a nanomanipulator (Omniprobe® and Easy Lift™), the taper-free InP TSL nanowires were pushed down to lie on the substrate for easy-lift operation. Then, they were mounted and aligned in the central gap of the PTP device to ensure that deformation was applied along the [111]B growth direction. Both ends of InP nanowires were firmly bonded on the PTP device across the PTP gap using platinum (Pt) deposition, as shown in Figures 2c-e. The whole welding process was carried out in a dual-beam FIB-FEG SEM (Helios NanoLab 600i, FEI, USA). In order to eliminate the $Ga^+$ irradiation on InP nanowires, only the segments used for bonding were exposed to the $Ga^+$ irradiation at a low current during the whole bonding process. More details about the tensile sample preparation can be found in Figure S1 (Supplementary material).

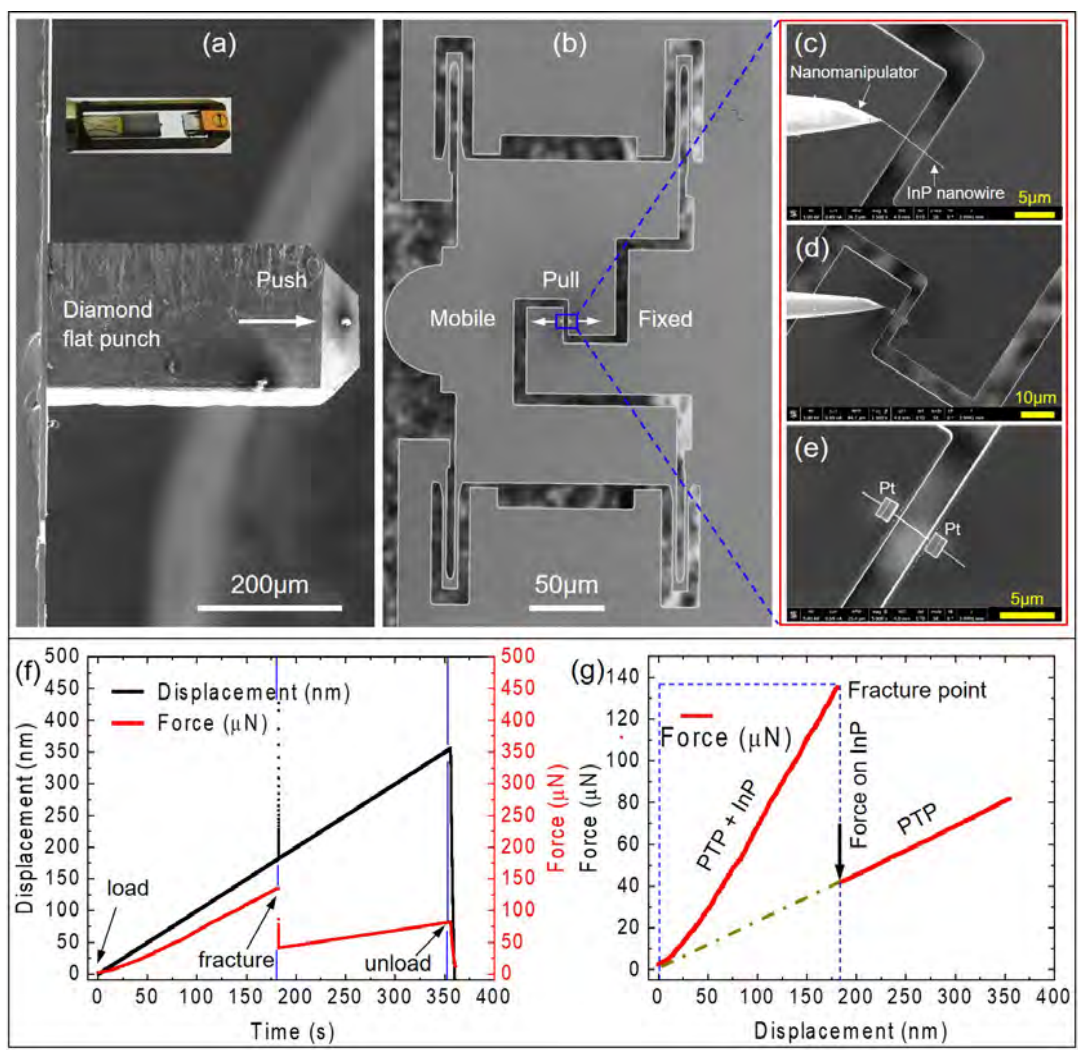

Figure 2. Experimental set-up for the *in situ* mechanical testing in a TEM. (a, b) SEM images of the PTP device and the flat punch. When the diamond flat punch pushes the semicircular end, the PTP device transforms this movement into a tensile opening of the central gap. (c-e) SEM images showing how to bond the InP superlattice nanowires to the PTP device across its gap using a nanomanipulator and a focused ion beam to weld the nanowire. (f) Displacement-force-time response of the InP TSL nanowire during the *in situ* TEM mechanical test. (g) Experimental approach showing how to calculate the net force applied on the InP TSL nanowires.

The displacement rate of the flat punch was 1 nm/s, leading to a strain rate of ~ 0.02 s$^{-1}$ in the deformed nanowires. The raw force and the displacement of flat punch were measured using the performech® Advanced Control Module that is coupled with the TriboScan™ software. Real-time videos were taken at 63 frames/s during the *in situ* TEM mechanical testing. The actual force applied on each InP nanowire was determined by subtracting the reaction force of PTP devices from the raw force, according to the design of the PTPT device [36] (see Figures 2f, g). The displacement of the gauge length was measured based on the images extracted from the real-time videos. Then, the engineering stress and strain could be calculated from this information and the nanowire diameter.

*2.3. Simulation techniques*

The electronic calculations in the present work were performed with the quantum espresso plane-wave pseudopotential code [44], using the density functional theory (DFT) within the generalized gradient approximation (GGA) [45] to the exchange-correlation potential and the projector-augmented wave method (PAW) [46]. The Kohn-Sham single particle wave functions were represented by plane-wave expansions with a cut-off of 52 Ry (707 eV). The sampling of the Brillouin zone was carried out by introducing a Monkhorst-Pack k-point grid [47] at a spacing of 0.03 Å$^{-1}$. The potential developed by Vashishta et al. [48] was used in the molecular dynamics (MD) simulations of the tensile deformation of the nanowires. They were carried out using the large-scale atomic/molecular massively parallel simulator (LAMMPS) [49]. The unit cell of InP zinc-blende nanowire contains four In and four P atoms in a face-centered cubic symmetry and the zinc-blende primitive unit cell obtained with DFT geometrical optimization was used as the basis of the MD model for InP TSL nanowires. The zinc-blende unit cell possesses a lattice constant $a$ = 5.836 Å. The nanowires were modeled with the x, y and z axes oriented along the [1$\bar{1}$0], [01$\bar{1}$] and [111] directions, respectively. Both untwinned and twinned nanowires were simulated and the distance between twin boundaries along the z axis was constant in the latter. The length of the nanowires $L$ was equal to 100 nm in all cases. Simulations of untwinned nanowires with diameters in the range $12 \leq D \leq 30$ nm were initially carried out to establish the minimum diameter to avoid any size effects. Afterwards, simulations of the InP TSL nanowires were performed for nanowires in which the ratio of the nanowire diameter to the twin boundary spacing $l$ was varied from $2 \leq D/l \leq 12$. In the case of $D/l$ = 12, the distance between twin boundaries was 2 nm and it was not possible to go to lower distances between twin boundaries (higher $D/l$) because the lattice structure of the nanowire would be modified if the distance between twin boundaries was below 2 nm. In the case $D/l$ = 2, the distance between twin boundaries in the model was 12 nm, very close to the experimental one (13 nm, Fig. 1e). Periodic boundary conditions were prescribed only along the z direction (tensile direction), while free surfaces were considered along the x and y directions.

The energy of the system was minimized using the conjugate gradient scheme at the beginning of each simulation followed by a dynamic stabilization of the system. To this end, an isothermal–isobaric (NPT) ensemble at atmospheric pressure and 300 K was performed for 100 ps. The number of time steps required for the simulation was chosen by trial and error to equilibrate various state variables. Once the system was equilibrated, the uniaxial strain was applied along the z axis ([111] direction) at a constant strain rate of $10^9$ s$^{-1}$, which has been successfully employed in previous studies [50, 51] taking into account the computational constraints of MD. The simulations of the tensile tests were conducted with the NPT ensemble using a Nose–Hoover thermostat. The prescribed displacement continued until the failure of the InP TSL nanowire. The time step prescribed in all the dynamic simulations was 1 fs. The interatomic stresses in our simulations were calculated using the Virial stress theorem [52].

**3. Experimental results and discussion**

*3.1. Materials microstructure*

The InP TSL nanowires grew along [111]B direction, perpendicular to the zinc-blende InP substrate (Figures 1a,b). Representative TEM micrographs of InP TSL nanowires at a lower magnification are shown in Figures 1c,d. The periodically twinned nanostructure formed along the whole length of the nanowires is similar to that reported in [8]. However, these as-grown nanowires are nearly taper-free with diameters of 210 ~ 250 nm, as opposed to the tapered nanowires in [8]. The average spacing between nanotwins is 13 nm, leading to a ratio between the nanowire diameter and the nanotwin spacing in the range 16 - 19. High-resolution TEM micrographs and the SAED pattern are shown in Figures 1e and f, respectively. After indexing the SAED pattern in Figure 1f, the $\Delta g$ ($\bar{1}\bar{1}1$) vector is exactly perpendicular to the twin interface in Figure 1e. It can be concluded that each InP TSL nanowire has a zinc-blende crystal structure with the [111] orientation parallel to the growth direction.

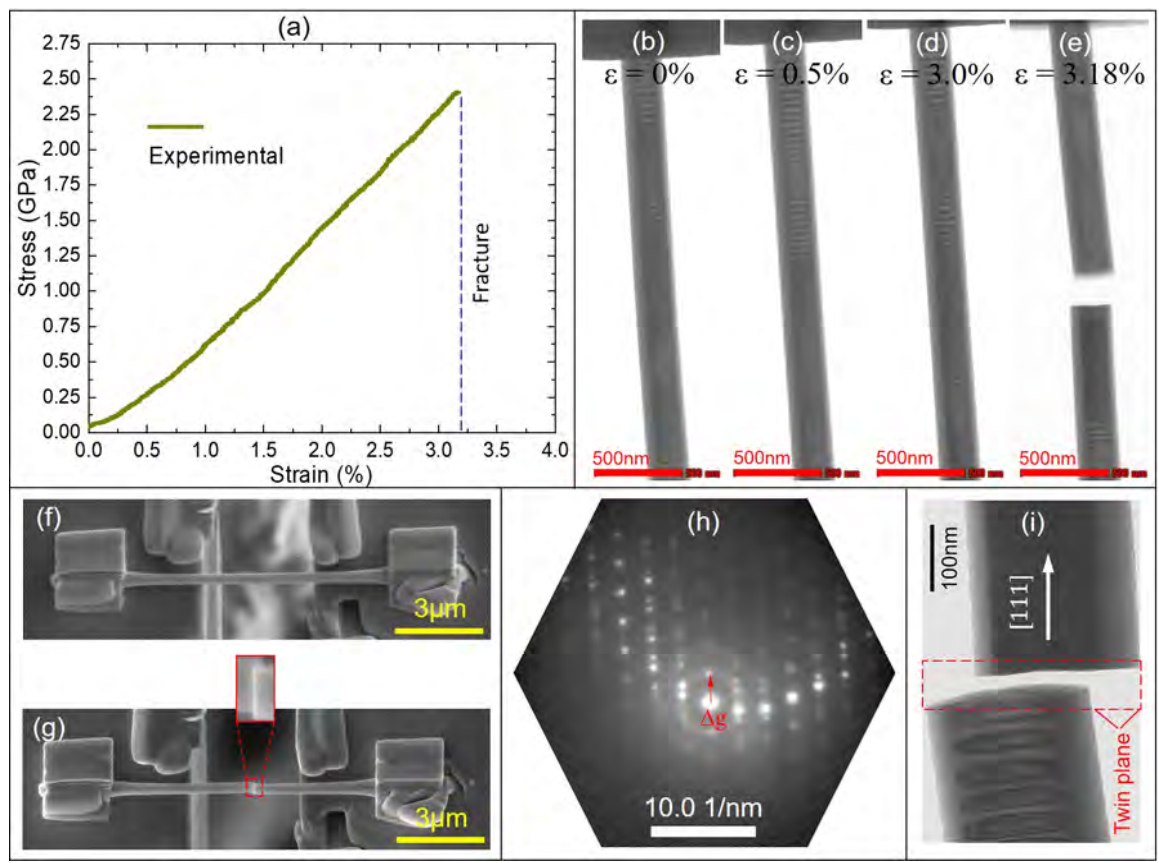

Figure 3. Tensile deformation of the InP TSL nanowire along the [111] orientation. (a) Representative stress *vs* strain curve of the InP TSL nanowire. (b-e) TEM micrographs of the nanowire at different applied strains (i.e. 0, 0.5, 3.0 and 3.18%). (f, g) SEM images of the InP nanowire before and after fracture. (h) SAED pattern showing the TEM beam direction is very close to the [111] zone axis, $\Delta g$ is nearly parallel to the tension direction. (i) Brittle fracture morphology with the crack plane nearly parallel to the twin boundary.

*3.2. In situ TEM mechanical deformation*

One representative stress-strain curve of the tensile deformation of the InP TSL nanowire is plotted in Figure 3a. The mechanical response was linear until fracture and the TEM images during deformation at different applied strains (0, 0.5, 3, and 3.18%) are shown in Figures 3b-e, respectively. They were extracted from Video S1 in the Supplementary material and show that the deformation was uniaxial without any traces of bending. This behaviour is different from the one obtained from compression-based deformation [23, 24, 34], in which bending is always present. The SEM micrographs of InP TSL nanowire before and after fracture are shown in Figures 3f and g, respectively. Figure 3h shows that the $\Delta g$ vector is nearly parallel to the tensile direction (or [111] growth direction). Fracture was brittle at a failure strain of 3.18% through a flat surface perpendicular to the growth direction (Figure 3i) and micrographs in Figures 3b to 3e do not show any evidence of inelastic deformation during in situ TEM tensile deformation even at very large elastic strains of 3%. Moreover, no evident defects were found along the nanowire. Figure 3h indicates that cracking occurred along the twinning interface, which seems to be the only defect in the crystalline nanowire. The relevant mechanical properties (fracture strain, fracture stress and elastic modulus) were obtained from a few successful tests in InP TSL nanowires along [111] orientation. They are depicted in Table 1 together with the nanowire diameter. In all cases, the tensile stress-strain curve was linear until brittle failure occurred along the twin interface perpendicular to the [111] orientation. The experimental elastic modulus (87 ± 17 GPa) is higher than that of pure zinc-blende InP nanowires, which is ~ 64 GPa for diameters between 35 and 40 nm [25].

The fracture strains were similar in all cases (2.9 ± 0.3%) with diameters in the range 210 to 250 nm. They are in good agreement with those reported in other semiconductor nanowires [40], but are much smaller than those in GaAs nanowires (10 - 11%) [24]. The tensile strength increased from 2.15 to 2.90 GPa as the nanowire diameter decreased from 250 to 210 nm but a size effect cannot be concluded from this limited data set. Size effects of the type 'smaller is stronger' have been reported in other nanoscale materials that were deformed by means of *in situ* tensile tests within a TEM, such as Si nanowires [53] and Mg single crystals [54]. The size effects in the latter were attributed to changes in the plastic deformation mechanisms, particularly when the size of the tensile specimen was below 150 nm, and similar conclusions were attained by means of MD simulations in coherent nanowires with tilted twins [55]. On the contrary, the deformation of Si nanowires was elastic until failure, and the fracture surfaces were flat and parallel to the (111) planes. The size effect in this case was attributed to the smaller probability of finding large surface defects as the nanowire diameter decreased. Similar 'smaller is stronger' size effect was reported in compression tests of GaN nanowires [20] but not in GaAs nanowires [24], where the failure strain was independent of the nanowire diameter (in the range 50 to 150 nm). Nevertheless, failure in compression took place by buckling of the nanowires, which is obviously influenced by the nanowire diameter and makes more difficult to identify a size effect.

Jiang et al. [56] fabricated arrays of nanotwinned Cu nanopillars that were vertically aligned along the $[1\bar{1}1]$ direction similar to the present InP TSL nanowires. They found that when the twin boundary was ≤ 2.8 nm, deformation involved plastic deformation by the nucleation of dislocations at twin boundaries followed by confined dislocation slip, leading to ductile fracture via shear localization and necking. On the contrary, nanopillars with a twin boundary spacing of 4.3 nm did not show any evidence of plastic deformation and failed in a brittle fashion from the twin boundaries. Local plasticity has been reported in semiconductor nanowires (i.e. GaN) deformed under compression [20]. Dislocations in this case were nucleated at the contact between the punch and the nanowire due to the stress concentration associated with nanoscale roughness and were piled up in this region until a local plastic zone was formed. In the case of the InP TSL nanowires tested under tension in this investigation, dislocation nucleation was not observed and the deformation was elastic until fracture. The twin spacing in the InP TSL nanowires was 13 nm and dislocation slip is very difficult to promote due to the directional covalent bonds in InP which limit the available slip systems and induce high Peierls stresses [24]. So, InP TSL nanowires failed in a brittle fashion from the twin boundaries which are higher energy surfaces. In addition, the zig-zag waviness in the nanowire surface (Figure 1e) leads to a stress concentration at the twin boundary which facilitated crack nucleation. It should also be noted that the deformation and fracture mechanisms may be different in nanowires tested under tension or compression. As indicated above, dislocation nucleation was observed at the contact between the punch and the GaN nanowire tested under compression, leading to local plasticity [20] but this mechanism is unlikely to develop under tension.  Chen et al. [23] performed *in situ* TEM compression tests in GaAs nanowires which contained periodic stacking faults (which are similar to the twin boundaries). The stacking faults prevented crack nucleation but failure in this case was triggered by the buckling of the nanowire.

Table 1. Mechanical properties of taper-free InP TSL nanowires along the [111] orientation.

| Diameter (nm) | Fracture strain (%) | Tensile strength (GPa) | Elastic modulus (GPa) |
|---|---|---|---|
| 250 | 2.76 | 2.15 | 77.9 |
| 234 | 3.18 | 2.40 | 78.4 |
| 210 | 2.79 | 2.90 | 104 |

*3.3. Comparison between simulations and experiments*

The elastic constants, stacking fault energy and lattice parameters of the InP TSL were obtained by DFT calculations. They are depicted in Table 2, together with those in the literature. They are in good agreement, particularly regarding the large anisotropy in the elastic moduli between [111] and [001] orientations. The optimised structure obtained from the DFT calculations was used to establish the models for MD simulations. Dos Santos et al. [57] reported that the elastic modulus of the untwinned [111] InP nanowire decreased with decreasing diameter when the diameter was smaller than 5 nm [57]. For this reason, the smallest diameter used in the present study was 24 nm. The stress-strain curves obtained from MD simulations of the untwinned InP nanowires with diameters in the range 12 nm to 30 nm at 300K are plotted in Fig. S2 (Supplementary material). The strength of the untwinned nanowires was ~7.3 GPa and independent of the nanowire diameter, while the elastic modulus increased from 82 to 85 GPa as the nanowire diameter increased. Failure of the untwinned nanowires was brittle and no energy dissipation mechanisms were found prior to failure. Simulations of InP TSL were carried out in nanowires with 24 nm in diameter and the twin spacing $l$ was changed from $2 \leq D/l \leq 12$. The corresponding stress-strain curves are plotted in Fig. S3 (Supplementary material), together with the one of the untwinned nanowire with the same diameter. They show that the strength of InP TSL (6.7 - 6.9 GPa) was smaller than that of the untwinned nanowire and that the influence of the twin spacing on the strength was minor when the distance between boundaries was longer than 3 nm ($D/l < 8$). Moreover, the presence of twins did not change the elastic modulus of the nanowire, which agrees well with the experimental results in Table 1 ($E_{[111]} = 87 \pm 17$ GPa) and slightly lower than the DFT predictions in Table 2.

Table 2. DFT results for the elastic constants ($C_{ij}$), bulk modulus (B), shear modulus (G), elastic modulus (E), stacking fault energy (SFE), Poisson's ratio ($v$), and lattice parameter ($a$) of InP TSL.

| Reference | $C_{11}$ (GPa) | $C_{12}$ (GPa) | $C_{44}$ (GPa) | B (GPa) | G (GPa) | $E^{[001]}$ (GPa) | $E^{[111]}$ (GPa) | SFE (mJ/m$^2$) | $v^{[001]}$ | $v^{[111]}$ | $a$ (Å) |
|---|---|---|---|---|---|---|---|---|---|---|---|
| This work | 99.9 | 56.1 | 43.8 | 70.7 | 33.2 | 59.5 | 113.9 | 15.2 | 0.36 | 0.25 | 5.836 |
| Ref. [57] | | | | 71.3 | | 56 | 106.4 | | 0.37 | 0.26 | 5.879 |
| Ref. [58] | 101.1 | 56.1 | 45.6 | | | 61.1 | 112.7 | | 0.36 | 0.24 | |
| Ref. [59] | | | | 72.4 | | | | | | | 5.869 |
| Ref. [59] | | | | | | | | 18 ± 3 | | | |

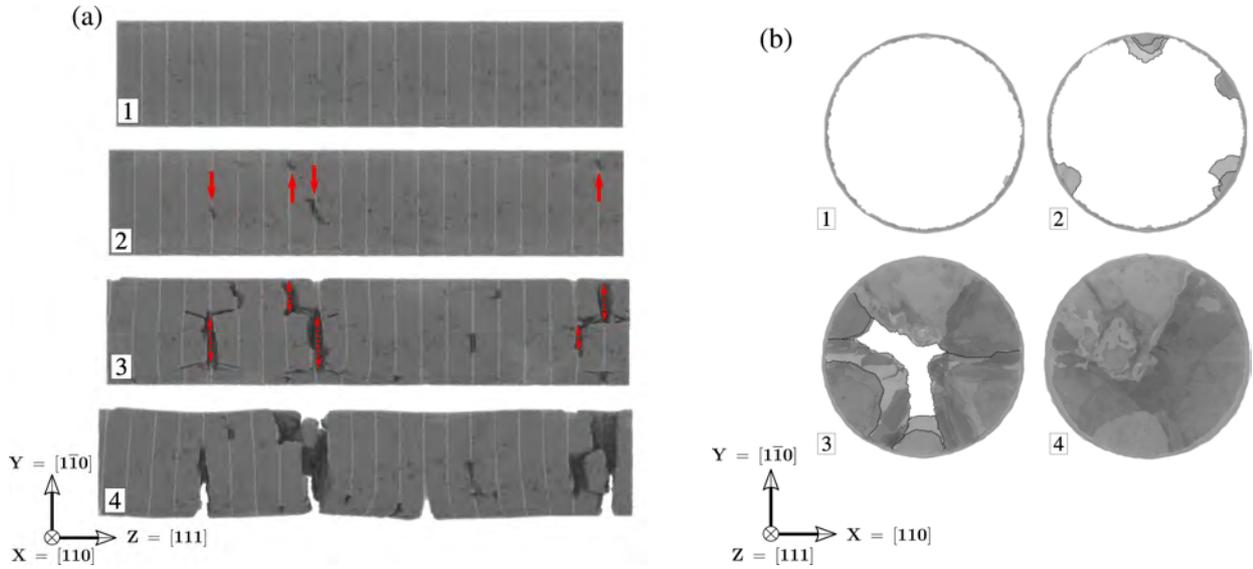

Figure 4. MD simulation results of crack initiation and propagation in an InP TSL nanowire with 24 nm in diameter during tensile deformation. (a) Section perpendicular to the twin plane. (b) Section parallel to the twin plane. The twin boundaries are shown as white lines in in (a). Crack initiation and propagation is evident at the twin boundaries (red arrows).

The stress-strain curves of the InP TSL nanowires obtained by MD did not show any evidence of nonlinear deformation mechanisms before failure. Fractures were nucleated at the intersection of the twin boundary with the nanowire surface and propagated along the twin boundary plane and towards the center of the nanowire (Figures 4a,b). The dynamic fracture behaviour simulated by MD can be found in Video S2 (Supplementary material). The fracture at the twin plane supports the experimental observations and indicates that the twin planes are partially responsible for the lower strength of the InP TSL as compared with the untwinned nanowires, although the differences are limited. Although fracture always took place along the twin boundaries, the mechanical strength of the InP TSL nanowires (2.15-2.90 GPa) was much lower than the strength obtained from the MD simulations (6.7-6.9 GPa). This discrepancy may be partially attributed to the extremely high strain rates of the MD simulations, but it is more likely that it was caused by the zig-zag waviness on the nanowire surfaces (Fig. 1e). The precise stress concentrations associated with these irregularities are difficult to estimate because they depend on the actual notch radius at the twin boundary (which is not known); however, it is obvious that they can have an important influence on the strength.

## 4. Conclusions

The tensile mechanical properties of taper-free InP twinning superlattice nanowires were determined by means of *in situ* mechanical tests in a TEM. The elastic modulus along the [111] orientation was 87 ± 17 GPa, while the fracture strain was 2.9 ± 0.3% close to the one reported in ZnO nanowires (~5%) but smaller than that of GaAs nanowires (10 - 11%). The tensile strength varied from 2.15 to 2.90 GPa for nanowires of diameter in the range 250 to 210 nm. Fracture was brittle in all cases and occurred by the propagation of a crack along the twin boundary interface. No evidence of inelastic deformation mechanism was observed neither in the experimental stress-strain curve nor in the TEM images before fracture. MD simulations of the tensile deformation of untwinned and twinned InP nanowires with different diameters (12 to 30 nm) at 300 K were carried out to assess the experimental data. The elastic modulus obtained from the simulations was fairly independent of the diameter and close to the experimental values. The failure strain and the tensile strength of the twinned nanowires in the MD simulations (~ 10% and ~ 6.7 GPa) were much higher than those observed experimentally but smaller than those simulated for untwinned nanowires, and these differences were attributed to the stress concentrations associated with the zig-zag waviness at the nanowire surface. Moreover, the simulations did not show any evidence of non-linear deformation mechanisms prior to fracture, that was triggered by the nucleation and propagation of a crack at the twin boundaries, in agreement with the experimental observations.

**Acknowledgements**

This investigation was supported by the European Research Council (ERC) under the European Union's Horizon 2020 research and innovation programme (Advanced Grant VIRMETAL, grant agreement No. 669141). Dr. Z. Liu would like to acknowledge the support from the Marie Sklodowska-Curie Individual Fellowship program through the project MINIMAL (grant agreement No. 749192). Dr. X. Yuan acknowledges the financial support from National Natural Science Foundation of China (No. 51702368), Hunan Provincial Natural Science Foundation of China (2018JJ3684) and Innovation-Driven Project of Central South University (2018CX045). The computer resources and the technical assistance provided by the Centro de Supercomputación y Visualización de Madrid (CeSViMa) are gratefully acknowledged. Additionally, the authors thankfully acknowledge the computer resources at Picasso and the technical support provided by Supercomputing and Bioinnovation Center of the University of Malaga and Barcelona Supercomputing Center (project QCM-2018-3-30). The Australian National Fabrication Facility is acknowledged for access to the epitaxial growth facility used in this work.

# SUPPLEMENTARY MATERIAL

# Mechanical behavior of InP twinning superlattice nanowires

Zhilin Liu[a,b], I. Papadimitriou[b], M. Castillo-Rodríguez[b], C. Wang[b], G. Esteban-Manzanares[b],

Xiaoming Yuan[c,*], H. H. Tan[d], J. M. Molina-Aldareguía[b], J. Llorca[b,e,*]

[a] *State Key Laboratory of High Performance Complex Manufacturing, College of Mechanical and Electrical Engineering, Central South University, 410083, P.R. China*

[b] *IMDEA Materials Institute, C/Eric Kandel 2, 28906, Getafe, Madrid, Spain*

[c] *Hunan Key Laboratory for Supermicrostructure and Ultrafast Process, School of Physics and Electronics, Central South University, Changsha, Hunan 410083, P. R. China*

[d] *Department of Electronic Materials Engineering, Research School of Physics and Engineering, The Australian National University, Canberra, ACT 0200, Australia*

[e] *Department of Materials Science, Polytechnic University of Madrid, E.T.S. de Ingenieros de Caminos, 28040 Madrid, Spain*

Corresponding authors: xiaoming.yuan@csu.edu.cn; javier.llorca@imdea.org

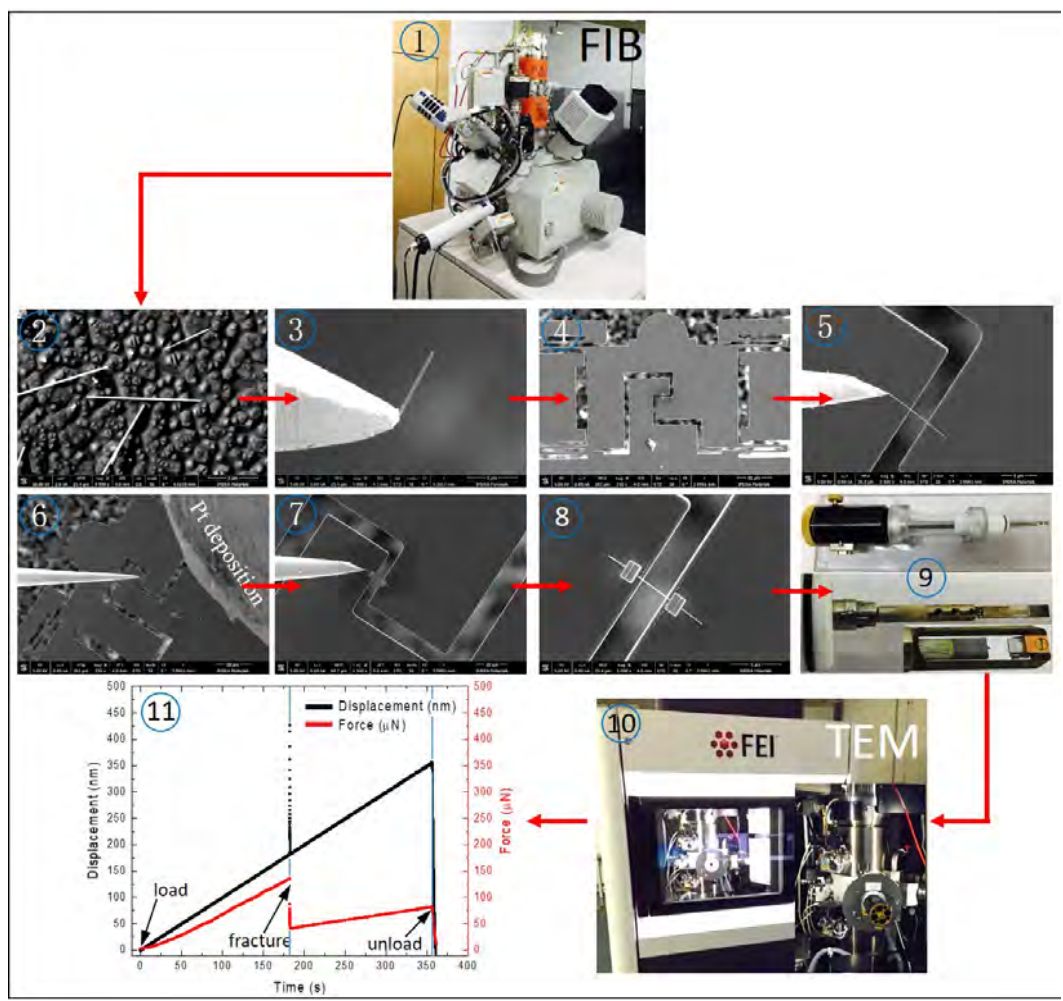

Fig. S1. 1 - 8: Tensile sample preparation of TSL InP nanowires, and 9 - 11: *In situ* mechanical testing of nanowires in a transmission electron microscope (TEM).

Video S1. The in-situ TEM mechanical testing of TSL InP nanowires.

Video S2. Molecular dynamics simulation of the tensile deformation and fracture of InP TSL nanowires.

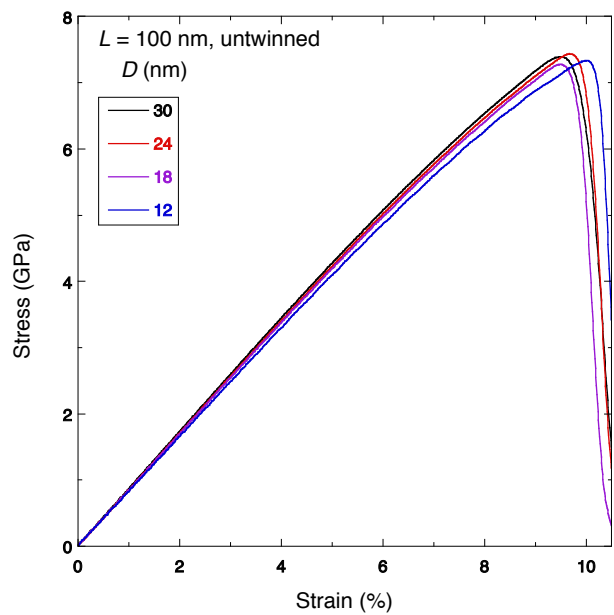

Figure S2. Molecular dynamics calculation results of the tensile stress-strain curves of the untwinned InP nanowires as a function of the nanowire diameter at 300K.

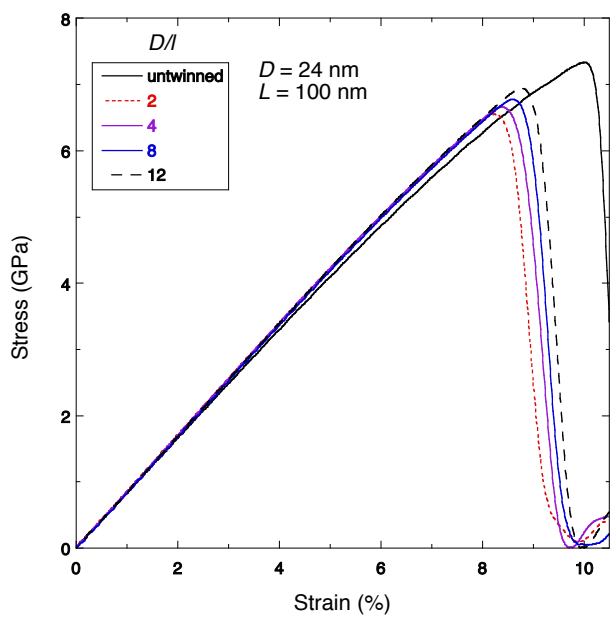

Figure S3. Molecular dynamics calculation results of the tensile stress-strain curves of the twinned InP nanowires as a function of the ratio between the diameter $D$ and the twin spacing $l$ at 300K